\documentclass[12pt,preprint]{aastex}

\shortauthors{Middleditch}
\shorttitle{GRBs, CC, Ia, Cosmology}
\font\eightrm=cmr8

\font\tworm=cmr2

\begin{document}

\title{Core-collapse, GRBs, Type Ia Supernovae, and Cosmology
}

\author{John Middleditch$^1$}

\altaffiltext{1}{Modeling, Algorithms, \& Informatics, CCS-3, MS B265,
                 Computer, Computational, and Statistical Sciences 
		 Division, Los Alamos National Laboratory, Los Alamos, 
		 NM 87545; jon@lanl.gov}

\email{jon@lanl.gov}

\begin{abstract}

The unexpected faintness of distant Type Ia Supernovae (SNe) has been 
used to argue for an accelerated expansion of the universe,
with the understanding that these are thermonuclear disruptions of 
accreting white dwarfs.  However, the high velocity and polarized 
features observed in SNe Ia, and their inverse relation to luminosity,
particularly for polarization, are consistent with an extreme version 
of the axisymmetry now seen in SN 1987A, which could be the result of 
double-degenerate merger-induced core-collapse.  This could be the 
correct paradigm for many Type I and II SNe, where Ia's are both 
thermonuclear {\it and} core-collapse objects, which leave weakly 
magnetized, rapidly spinning ($\sim$2 ms) pulsars.  In this paradigm, 
a Ia/c is produced from the merger of two degenerate cores of common 
envelope Wolf-Rayet stars, or of two CO white dwarfs.  Thus the same 
explosive mechanism that underlay 10--15 M$_\odot$ in SN 1987A, underlies 
only 0--1.5 M$_\odot$ in SNe Ia/c.  Its now visible polar blowout features 
produce the observed high velocity and polarized spectral features in Ia's, 
and its equatorial bulge is much brighter in Ia's, due to the greater 
fraction of $^{56}$Ni contained within it.  Such merger SNe become 
classified as Ia's when viewed from the merger equator, and Ic's when 
viewed from the poles (given sufficient matter in excess of the 1.4 
M$_{\odot}$ lost to core-collapse), where a hypernova signature and a 
gamma-ray burst (GRB) will be observed for lines of sight close to the 
merger axis.  This complication may mean that cosmology determined strictly 
from Ia's alone is flawed, because the local sample may be selectively 
biased.  Finally, all GRBs start as the short-duration, hard-spectrum 
variety (sGRBs), after which some are modified to the long duration/soft 
spectrum variety ($\ell$GRBs) by interaction with the merger common 
envelope and/or previous polar ejection.  Thus the initial photon spectrum 
of nearly all GRBs is known.

\end{abstract}


\section{Introduction}
Type Ia supernovae (SNe Ia) have been used by at least two groups, and 
all without any explicit foreknowledge of their progenitors,\footnote{In 
a Las Campanas 2.5-m run on SN 1987A during 1995 Feb.~my late colleague, 
Dr.~Jerry Kristian, mentioned to me that he considered the Ia cosmology 
effort ``a perversion of the science,'' because no one knew what SNe Ia 
really were.} to argue that the expansion of the universe is accelerating, 
and hence for the existence of ``dark energy,'' or a cosmological constant, 
$\Lambda$ (see, e.g., Riess et al.~1998; Perlmutter et al.~1999).
This has the appearance of 
convenience, as it helps several other lines of inquiry, including the 
scale size of the fluctuations of the surface of last scattering of the 
cosmic microwave background (CMB), and measurements of the clustering 
mass on large scales (see, e.g. Eisenstein et al.~2005), converge to a 
consistent set of parameters, generally $\Omega_m \sim 0.3$ and 
$\Omega_\Lambda \sim 0.7$.  
However, at present SNe Ia represent the only firm, direct evidence for 
the existence of dark energy (see, e.g., Conley et al.~2006).

If dark energy has been convenient for cosmologists, it is certainly
not so for the Standard Model, which allows no such subtle effect by
120 orders of magnitude (Weinberg 1989).  
Still, a lot of other recent efforts have gone into shoring up the case 
for cosmic acceleration (see, e.g., Clocchiatti et al.~2006), and there is 
no doubt that all are seeing the same effect.  
In addition, many have pointed out that the effect is still present
even without using the width-luminosity (WL) relation, determined for 
local SNe Ia a decade earlier by Phillips and others (see, e.g., Phillips 
et al.~1999, and references therein), to correct the Ia luminosities.
However, recent observations have shown that SNe Ia have high velocity 
and polarized line features (HVFs \& PLFs) at the few \% level 
\citep{Le05,Maz05a}, as well as a few tenths of a \% mean continuum 
polarization. 
In addition, core-collapse (CC) in Ia's may be {\it necessary} to explain 
many of the ms pulsars (MSPs) in the globular clusters (GCs -- see 
$\S$\ref{sec:Ia/Ic?}), and the abundance of Zn \citep{Ko06}.

In this letter I argue that many SNe Ia are caused by double-degenerate
merger-induced CC (DD) which are classified as Ia/c's
when viewed from the merger equator/poles, the latter occurring only when
sufficient matter exists in excess of that lost to the CC to screen
the Ia thermonuclear (TN) products.  This complicates the use
of SNe Ia in cosmology.  I also argue that nearly all gamma-ray bursts
(GRBs) are due to DD, with all starting as the short duration and hard 
spectrum variety (sGRBs), and only later are some modified to the long 
duration/soft spectrum variety ($\ell$GRBs), by interaction with the 
common envelope (CE) and previous polar ejection.

\section{What are SNe Ia/Ic?}
\label{sec:Ia/Ic?}

In spite of initial mass estimates as high as 40 M$_{\odot}$ for
some SNe Ic progenitors (SN 2002ap, 2003dh -- Mazzali et al.~2002, 
2003), no Ia or Ic progenitor has ever been identified 
(see, e.g., Maund et al.~2005), 
and much lower mass binaries,
such as CE Wolf-Rayet (WR) stars, (see, e.g., DeMarco 
et al.~2003; Howell et al.~2001; and the data in G\'orny \& Tylenda 
2000), have never been eliminated as Ia/c progenitors (Gal-Yam et al.~2005).
The ``usual'' Ia paradigm, (gradual) accretion-induced collapse of a 
white dwarf (WD), or single degenerate (SD), requires ignition and burn
without core-collapse at the Chadrasekhar mass (1.4 M$_\odot$), and 
also has difficulty explaining the absence of H and He from the WD 
and/or advected from the mass-donating companion star \citep{Mat05}, 
the unsuitability of cataclysmic variables as progenitors \citep{SB05},
the four Ia's in the last 26 years of the colliding elliptical/spiral
galaxies of NGC 1316 
(see, e.g., Tsvetkov et al.~2005, and references therein),
and Ia's which
produce much more than 0.1 M$_\odot$ of $^{56}$Ni from a single ignition 
source (Brown et al.~2005), though multiple sources of ignition have been 
discussed by R$\ddot{\rm o}$pke et al.~(2006).  Finally, in addition to 
their HVFs and PLFs, Ia's also show a broad range of diversity in their 
velocity gradients and correlations with Si line ratios and 
$\Delta m_{15}(B)$ (Hachinger et al.~2006; James et 
al.~2006), all unlikely side effects of simple TN disruption.

The morphology of the explosion of SN 1987A has now been clear for
a number of years (NASA et al.~2003; Wang et al.~2002; Middleditch 2004, 
hereafter M04).  
A polar blowout feature (PBF) approaches at about 50$^{\circ}$ off our 
line of sight (see, e.g., Sugerman et al.~2005).  It partially obscures 
an equatorial bulge/ball (EB), behind which a part of the opposite, 
receding PBF is visible.  The PBFs and EB are approximately equally 
bright.

SN 1987A is thought to have ejected about 10--15 M$_{\odot}$ (see,
e.g., Woosley 1988),\footnote{All such models are, of course,
invalid for DD due to the differences 
in mixing and the CC process.} and, because of
the blue supergiant nature of the progenitor \citep{Sa69}, the rings 
\citep{Bu95}, the ``mystery spot'' 
\citep{Me87,Ni87},\footnote{This feature still contained
10$^{49}$ ergs some {\it six weeks after} being hit with a relativistic
jet/beam (M04 -- the deviation off the line of sight taking only
8 extra days).  In this interpretation, the ``mystery spot'' 
corresponds to the GRB ``afterglow.''}
the mixing \citep{Ma87,Co88}, the polarization (Schwarz \& Mundt 1987; 
Barrett 1988), and the possible 2 ms pulsar remnant 
\citep{M00b},\footnote{On
two occasions shortly before his death, I discussed the pulsar
search effort on SN 1987A with Kristian, and the only thing both 
of us could conclude was that the 2.14 ms signal was very likely 
real.  This is still the case.} is very likely to have been the 
result of a DD merger of two stellar cores.  
Occam's Razor alone would argue that Ia's and Ic's are
the result of the same process(es), but there are many
other good reasons.  If Ia's and Ic's are the result of 
common-envelope WR DD mergers
or CO-CO WD mergers common in GCs (which leave MSPs -- see, e.g., 
Chen et al.~1993), then the limit for mass ejected is 
near 1.5 M$_{\odot}$, consistent with the $\sim$1 M$_{\odot}$ limit 
of $^{56}$Ni produced in SNe Ia \citep{Ho06}.  

If SNe Ia/c are the result of the same explosive process
that underlay 10--15 M$_{\odot}$ in SN 1987A (CC), but 
which instead only underlies 0.5 M$_{\odot}$, the outcome will
be even more extreme than the geometry of the SN 1987A remnant.
The PBFs will have higher velocities, and the equatorial/TN
ball (TNB) will be much brighter due to the
greater concentration of $^{56}$Ni, but its expansion
velocity will not necessarily be higher than in IIs, as the
mean Z for the outer ejecta will be higher.
Thus, there is no need to 
invoke exotic mechanisms such as ``gravitationally confined 
detonation'' to explain SNe Ia (Plewa et al.~2004).  
When viewed close to the poles of 
the merger, one of the two PBFs will obscure the TNB (if there is
sufficient mass above the merger pole(s)), and show lines 
of r-process intermediate mass elements (IMEs) from its end, and the 
SN will be classified as a Ic, and for views very close to the poles,
a hypernova signature will be seen, in addition to a 
GRB (M04).\footnote{Thus it should not be surprising that the 
faint GRB 031203 was associated with the bright SN Ic 2003bw 
\citep{Mal04}, but the other way around {\it would} be, as long as 
there were no late rebrightening, as was the case for GRB 050525A 
and SN 2005nc \citep{DV06}.}  

WR and CO-CO mergers could account for 
GRBs in both young and old populations (I and 
II -- see, e.g., Mannucci et al.~2006).  And although 
the Ia:Ic ratio in elliptical galaxies is $>$30
(see, e.g., the tables in van den Berg et al.~2005),
matter in excess of 1.4 M$_{\odot}$ from two WDs in 
older populations (II) is rare.  This picture of ``leaner'' mergers in 
non-actively star-forming galaxies is consistent with Hamuy et al.~(2000),
Sullivan et al.~(2005), and Wang et al.~(2006a), and may mean that the 
PBFs run out of sufficient matter to obscure the TNBs before these, in 
turn, run out of ejected TN ash, leading to relatively more Ia's.
Because only sGRBs have ever been seen in elliptical galaxies, and
globular clusters (GCs) require DD events to make many of their 
MSPs (with ``recycled'' MSPs restricted mostly to core-collapsed clusters 
[CCd] such as Ter 5),\footnote{Ter 5's lower luminosity/duty-cycle (the latter 
so much so as to overcome the effects of the former) MSPs \citep{Ra06}, and 
many of those of M15 are consistent with this picture, but others and all of 
those in non-CCd clusters such as 47 Tuc are not (Chen \& Ruderman 1993; M04).}
four remarkable, fairly robust conclusions can be drawn.  First,
many SNe Ia in older populations {\it must} in fact be such CC events.
Second, because DD events must occur in elliptical galaxies and 
numerically exceed by far NS-NS merger events, the vast majority of the 
sGRBs in ellipticals result from their lean, DD events.  Third, {\it all} 
GRBs start as sGRBs, and some convert to $\ell$GRBs by impacting the CE, 
and some are further modified by impacting previous polar ejecta.
Fourth and finally, sGRBs can be produced by DD events without 
distantly observable afterglows or SNe \citep{GY6,Fyb6,MDV6,Gh6}.

Unless the view {\it is} very near polar, this geometry will have 
no difficulty in producing split emission line(s) on rare
occasions, as was seen in SN 2003jd, and thus again there is
no need to invoke other explosion mechanisms 
\citep{Maz05b}.\footnote{And
there is absolutely no need to invent an entire population (III) to 
account for GRBs (Conselice et al.~2005; M04).}
When viewed sufficiently far from the poles, the TNB will dominate
the luminosity, S/Si absorption lines will appear/deepen, the SN will 
be classified as a Ia, but similar high velocity lines will 
also appear from the sides of the PBFs due to material advected 
from the TNB, in addition to IME lines, including Si again.  The 
assumption implicit in Ia cosmology is that the TNB will be a 
standard candle \citep{PE01}, which can be compared to the 
redshift of the host galaxy to determine the expansion 
properties of the universe.  


\section{The Evidence}
\label{sec:Evidence}

\subsection{The Ubiquitous High Velocity Features in Ia's}
\label{sec:HVFIa}

In the DD paradigm these \citep{Maz05a} result from 
SNe Ia having two PBFs of finite angular width.  The
half width angle of the PBFs can be estimated from the ratio of
the numbers of nearby SNe from catalogs.  If we
let $R_1$ be the TNB radius, below which sufficient visible EW in 
Si and
Fe lines would result in the SN Ia/c being classified as a Ia, $R_2$ 
the radius of the limit cones of the PBFs at the same given time, 
$\theta$ the half angle of the PBF cones, $\zeta$ the half angle of 
the cone circumscribed around the (assumed spherical) TNB and 
containing the circular boundary of the base of one PBF limit cone 
(the visibility divider between Ia's and Ic's), and $\phi = sin^{-1} 
({R_1}{R_2}^{-1})$, then $\theta = \zeta + \phi$.  

Counting SNe with radial velocities $< 10,000$ km s$^{-1}$, from SN 
1995O to 2004cg in the SAI catalog \citep{Tsv05},
gives 282 Ia's, 66 Ic's, and 301 Type II SNe,\footnote{Excluding
Type IIn's.} for a Ia:Ic ratio of 4.27:1, consistent with a 2:1 Ia:Ibc 
ratio in spirals from Cappellaro et al.~(1997).
Using a 1:5 ratio for $R_1$ vs $R_2$ (which 
might be an overestimate) gives $1 - cos{\zeta} = {5.27}^{-1}$, or 
$\zeta = 35.9^{\circ}, \phi = 11.5^{\circ},$ and $\theta = 
47.4^{\circ}$, a value consistent with the image of SN 1987A, and
big enough so that HVFs from the PBFs will almost always occur, 
no matter what the orientation from which SNe (classified as Ia) 
are viewed.  

Benetti et al.~(2005; hereafter B05) have divided SNe Ia into three 
categories, FAINT, and the brighter high and low velocity gradient 
(HVG and LVG) SNe.  In the PBF/TNB paradigm, FAINT Ia's are indeed 
intrinsically faint, producing little $^{56}$Ni, and their high 
velocity gradients and $\sim$9,000 km s$^{-1}$ expansion velocities 
are the result of only a small amount of material in their ejecta, 
and thin polar ejecta and/or a near-equatorial view.  The HVG SNe 
Ia are brighter, due to more $^{56}$Ni, and have $\sim$12,000 km
s$^{-1}$ expansion velocities, but are viewed substantially off 
the DD merger equator, where the opacities of conal PBF and 
advected TNB material diminish rapidly with time.  The LVG SNe Ia 
are also bright, but have a low velocity gradient and $\sim$10,000 
km s$^{-1}$ expansion velocities because they are observed from 
near the equator of the DD merger.  The increasing/decreasing 
evolution of $R${\tworm ~}(Si{\eightrm ~II}) (i.e., temperature) 
with time of HVG/LVG SNe Ia is consistent with this interpretation 
(Figure 2 of B05).

Four out of five of the FAINT Ia's in B05
fall below the WL relation by 1--2 {\it whole} magnitudes, 
and yet their $\Delta m_{15}(B)$'s fall just barely
outside the validity range for the WL relation ($<$1.75).
However, the spectra of all of these show a deep TiII trough
between 4,000 and 4,500 $\AA$, visible even at cosmological
distances, and thus can be safely excluded from the sample 
used for cosmology.  But problems could still arise if a 
continuous class exists between FAINT and HVG and/or LVG,
as might reasonably be expected.

In this case, it might be possible that, because of the assumption 
of an invalid paradigm for SNe Ia (SD), and the desire to avoid 
contaminating the sample of Ia's by including SNe which appeared 
too ``Ic-ish,'' with too much EW in IME lines, and too little in 
the SII and SiII lines, a local sample of Ia's was selected 
in which many were viewed very close to the equator of the
DD merger.  When the high velocity of the small amount of 
matter in near-polar ejecta of Ia's exposes a fraction of the
TNB to non-equatorial views during the interval when $\Delta m_{15}$ 
is measured, insufficient WL corrections could result in distant 
Ia's which appear to faint for their redshifts.  

\subsection{Polarization}
\label{Polar}

Continuum polarization near the 0.9--0.4\% level was detected in the Type 
IIP SN 1987A \citep{Sc87,Ba88}, two out of three Type IIs observed by Wang 
et al.~(1996), 1994Y, and 1995H, at nearly {\it twice} that level in the 
Type IIb SN 1993J \citep{Tr93},\footnote{In the case 
of SN 1993J, the PBFs are, and/or appear longer due to the low amount of H,
and/or a near equatorial view, or both.  Thus 1993J is a ``missing link''
between Type II and Type Iabc SNe.} and 0.2--0.5\% in the Type IIP 
SN 1999em, 7 to 159 days after maximum, by Leonard et al.~(2001).
Thus, continuum polarization in IIs is (and should be) common due to 
their PBFs having approximately equal brightness as their EBs.  

Continuum polarization is markedly low or absent in Ia's, typically
$<$0.2\%, with the exceptions of SN 1996X at 0.3\% (Wang et al.~1997),
the FAINT SN 1999by at 0.3--0.8\% 
\citep{Ho01}, the LVG SN 2001el, near 0.2--0.3\% \citep{Ka03}, 
the LVG SN 2003du at 0.3\% \citep{Le05}, and SN 2005hk at 0.4\% 
\citep{Ch05}.  This is exactly what is expected for Ia's because 
the TNB is roughly spherical, and dominates the luminosity of Ia's,  
where the PBFs are viewed from the sides, but are not bright enough to 
matter.  SN 1999by was highly polarized because it was a faint SN Ia, 
and the PBFs did matter.  SN 2005hk was less luminous (likely a member 
of the FAINT class) than 2001el, and thus its level of continuum 
polarization is higher.

In contrast, the Ca II 800 nm IR triplet and the Si II $\lambda$ 
6355 $\AA$ (absorption) lines in nearly all SNe Ia 
show higher polarizations, particularly those with high velocities, 
which is a result of the lines preferentially originating from the 
sides of the PBFs, visible (by definition) in all Ia's
(Kasen et al.~[2003] on SN 2001el
``The high velocity triplet absorption [800nm Ca II] is highly polarized 
$.~.~.~$''; Wang et al.~[2003] ``$.~.~.$ is distinct in velocity 
space from the photospheric Ca II IR triplet and has a significantly higher 
degree of polarization ($\approx$0.7\%) and different polarization angle 
than the continuum $.~.~.$ kinematically distinct feature with matter 
distributed in a filament $.~.~.$ almost edge-on to the line of sight 
$.~.~.~$'').

Line polarization occurs in increasing strength in overluminous, normal, 
subluminous, and high velocity SNe (HV -- broad lines, blueshifted 
up to 15,000 km s$^{-1}$) in observations \citep{Le05}, as 
well as naturally in the TNB/PBF paradigm.  Clearly the
polarization of Ia's should increase as apparent luminosity (the visible 
TNB) decreases,  until so little matter remains to be ejected into the 
PBF that it becomes optically thin (see $\S$\ref{sec:Ia/Ic?}).  Thus
the strength of the line polarization is only 0.2\% in the overluminous SN 
2003du, but is 2\% in lower luminosity HV SNe such as 2004dt \citep{Wa04}, 
again consistent with the PBF/TNB paradigm for Ia/c's.\footnote{Kasen 
and Plewa (2005) interpret the features of 2001el through gravitationally 
confined deflagration-induced detonation.  However, this mechanism is 
unlikely to produce the inverse relation between polarization and Ia 
luminosity described above, although Wang et al.~(2006b), having 
communicated about this work on 2006, August 21, explore the relation 
for SiII in greater detail (on 2006, Nov.~29, and without citing this 
work!), and surmise ``strong'' support for delayed detonation (see, e.g.,
Khokhlov 1991).}

Polarization in Ic's is more complicated, as those seen most end-on
will also have blueshifted HV lines, but will appear to be spherical, 
so little polarization will result.  
However, when both PBFs are visible, the extension in one direction 
will produce polarization, just as it does in Type IIs,
and may exceed that in IIs due to the rapid extension of the higher 
velocity PBFs and the obscured, or even the non-obscured, but faded, 
EB/TNB.  These are the Ic analog of the HV Ia's.  The continuum 
polarization of the Ic SN 2002ap rose from 0 to 0.5\% from -6 to +3 
days from maximum \citep{Wa03}, and later to 1.0\% and 1.6\% on days 
16 and 37 respectively \citep{Le02}.

\subsection{The Merger Paradigm and the Luminosities of SNe}
\label{MoreM}

The merger paradigm also explains why Types Ib and II SNe produce 
relatively little $^{56}$Ni, because their C and O layers are 
diluted with He, in addition to H for IIs, due to the merger 
process.  Aside from Type II-L SNe, Type IIs, initiated through 
Fe photodissociation catastrophe, suspected for SN 1986J (Bietenholz 
et al.~2004), may so far be the only known exceptions to 
the merger paradigm, and the vast bulk of these, i.e., those which do 
not continue to collapse into black holes, should be brighter than DD 
Type IIs due to their C and O layers remaining relatively 
undiluted at the time of CC, in addition to their embedded, 
strongly magnetized pulsars.  In DD Ia/c, which all
lack H and He, not only are the C and O layers undiluted, they are 
intimately mixed.  So if IIs manage to produce {\it some} $^{56}$Ni,
then even CC Ia's ought to produce much {\it more}.

\subsection{Conclusion}
\label{sec:Conclusion}

I have argued above that many SNe Ia and Ic are the result of the 
same process, DD events of WR or CO-CO WD binaries, in order to 
produce their HVFs and PLFs, and their inverse relation to Ia 
luminosity, many of the MSPs in old populations, sGRBs and $\ell$GRBs, 
the four recent Ia's of the colliding spiral/elliptical galaxies 
of NGC 1316, the $>$1 M$_{\odot}$ of $^{56}$Ni produced in 
SN 2003fg, and possibly the observed abundance of Zn.  This 
complicates the utility of Ia's to cosmological efforts, 
with at least one possible systematic effect which could produce 
distant Ia's which appeared to be too faint for their redshifts
(see the end of $\S$\ref{sec:HVFIa}), potentially adding to other
systematics, some of which could have the opposite effect,
such as Malmquist bias.  In addition, a 7\% Ibc contamination 
level is sufficient to produce $\Omega_{\Lambda} = 0.7$ from no 
effect (Homeier 2005).  Small wonder then that statistical 
considerations alone may rule out {\it any} cosmology derived 
from SNe Ia (Vishwakarma 2005).  Is there likely still {\it any} 
$\Omega_{\Lambda}$ effect at all left in Ia cosmology?~ Only time 
will tell.

I have also argued above that all GRBs begin as sGRBs before
some are changed into $\ell$GRBs by interaction with matter
in the binary merger CE and previous polar ejection.  This
could help to greatly simplify the study of GRBs because
the initial photon energy spectrum of the vast majority
of GRBs is now more-or-less known.

Using astronomical sources of any kind to determine the nature of
the universe is a tough business.  Divining the structure of SNe using
only spectroscopic data (and even the occasional polarization spectrum), 
an inverse problem for which very little progress has been made over
several past decades, only increases the difficulty immeasurably.  In
SN 1987A nature has revealed the origin and structure of most SNe, including
SN 1993J, and all we have to do is to pay attention.  The SN Ia cosmology 
effort sets all records for blood\footnote{We're all a decade older.} and 
treasure spent on a problem in astronomy.  It was a gamble, but one that 
may not pan out.  The attempt to use a misunderstood phenomenon 
to measure cosmological parameters is perhaps understandable, 
but ultimately unwise.

Supernovae are indeed wondrous objects, with up to 99\% resulting
from DD events.  These may all produce $\sim$2 ms 
pulsars which, in their first few seconds after birth, 
may very well be the only frequently detectable gravitational 
radiation sources, and
which, for at least a few years, shine in the optical band \citep{M00b}.
Many of their progenitors are not massive stars, and many produce jets, 
GRBs, and what astronomers have dubbed ``mystery spots,'' ``GRB afterglows,''
``Type II-P plateaus,''
``hypernovae,'' ``collapsars,'' ``supranovae,'' and my personal favorite, 
``The Beam from Hell,'' from what could appear to be unremarkable 20th 
magnitude blue stragglers, which can incinerate half the planet from a great 
distance with little or no warning.  What they are {\it not}, however, 
are easily utilized standard candles, and $\Omega_\Lambda = 0.7$ may
{\it not} be the ``correct'' answer.  It may be impossible to get a 
``clean'' sample of SNe Ia \citep{Be04,Bl05}.

The attempt to use SNe IIP to measure cosmological parameters 
(see, e.g., H$\ddot{\rm o}$flich et al.~2001; Nugent et al.~2006), 
faces the difficulty that even though most of these result from DD 
mergers of more massive stars, then, like SN 1987A (and possibly 1993J), 
they will still manage to produce beams/jets which may or may not impact 
their previous polar ejection to produce a ``mystery spot,'' (or two).

\acknowledgments I have had the time to write this course correction for 
science only because CCS-3 has supported me during this interval when I was 
without funding, and for that I am extremely grateful.  I would like to 
thank Drs.~Geoffrey Burbidge, Falk Herwig, Peter Nugent, and an anonymous 
referee for useful suggestions which helped me to improve this manuscript.  
I would also like to thank Jerry Jensen for conversations and bringing 
this issue to my attention.  This research was performed under the 
auspices of the Department of Energy.

\section{Appendix -- The Anatomy of a Rejection}

Dear John:

Title:  Core-Collapse, GRBs, Type Ia Supernovae, and Cosmology
Authors:  John Middleditch
 
Your revised manuscript was sent to a new expert referee.  This
person was chosen as much as possible to have no research or
personal ties to the first referees and thus could provide an
independent view of your paper.  The new referee has returned a
report that is appended below.  Unfortunately, this referee
comes to the same conclusion as did the first two referees.
The present reviewer finds substantial problems in your work,
and does not believe that the paper can be repaired.  He/she
recommends rejection of the paper, as did the previous referees.
Therefore I regret to tell you that the Astrophysical Journal
Letters will not be able to publish this manuscript.  We will
be happy to consider new manuscript submissions from you in the
future.
 
Regards,
Chris Sneden, Letters Editor
THE ASTROPHYSICAL JOURNAL LETTERS
apjletters@letters.as.utexas.edu
 
        Dear Chris,
 
        I actually missed the rejection msg until yesterday!
Saw the title but thought it was more apology fluff related
to the other msg on the same day (2006 Dec. 19).
                                                                               
        Anyway, the following is an FYI for the record.
 
        Do with it what you will.
 
        I am severely disappointed that ApJL does not see it
as their duty to publish legitimate scientific dissent,
particularly against a large group of astronomers who have
gotten more paranoid about such dissent, particularly in light
of two very recent SN developments, than Richard M. Nixon
just prior to his resignation.
 
        -John Middleditch
 
On Tue, 2006-12-19 at 14:41, apjletters@letters.as.utexas.edu wrote:

{\it This paper should not be published and the debate should
end.}
 
{\bf Four} Type Ia SNe in NGC 1316 within 26 short years, 1.2 solar
masses of $^{56}$Ni from 2003fg, and the debate should end??  That's
a version of Bob Kirshner's reply on 2005, Aug. 9: ``Well gee, we
got the right answer!''

{\it I am sure the author will complain that this is just the
establishment resisting new ideas. It isn't. Current
paradigms, which are in fact continuously evolving, are a
consequence of decades of observations and theory by the
community, no one or small group of people. They can be
overturned, one by one, by hard work and compelling
arguments, but not in one broad swath based chiefly upon
opinion and not calculation.}
 
Putting tons of publications, numbers thrown at a problem,
especially calculations, even {\bf decades} of work by many people,
{\bf above} observational fact, strikes me as staggering arrogance.
Doing so while at the same time ignoring perfectly valid
observations of others (see below re: 2.14 ms pulsations from
SN 1987A), strikes me as hypocrisy, never mind that in one
instance the vulnerability is systematics while in the other
it is the absence of an individual overwhelmingly significant 
result.  It doesn't matter how many people have published on
Ia cosmology for how long.  A systematic exclusion of certain 
types of Ia's, which ought to exist, from the local sample will
take down {\bf everyone's} results, one of the points made in
this paper.  Observational fact trumps everything else.  Calculations 
will never fail to err without {\bf rigorous} observational constraint.  
ApJL is {\bf intended} as a mechanism to overturn or reinterpret just 
such decades of hard work/science, using just such cogent arguments:
 
"Timeliness -- A Letter should have a significant immediate
impact on the research of a number of other investigators or
be of special current interest in astrophysics. ... A Letter
can be more speculative and less rigorous than an article for
Part 1 but should meet the same high standard of quality."
 
{\it The author attempts a grand synthesis that brings together
aspects of SN 1987A, Type Ia supernovae, long gamma-ray
bursts, short gamma-ray bursts, and core collapse supernovae
of all kinds. At the heart of this is a merger model -
either of two WR stars or two white dwarfs.}

{\it He argues - or states - among other things that SN Ia's 
are WD mergers accompanied by core collapse. He is probably
right that such a merger would lead to neutron star
formation if over 1.4 solar masses stayed bound, but quite
wrong about the Ni mass. Spectra and light curves demand
about 0.7 to 0.8 Msun plus a few tenths solar mass of Si
through sulfur. The result of the merger of two white dwarfs
would not give enough mass in close proximity to the
collapse to make these elements. Please present or reference
a calculation that proves otherwise.}

Given the circumstances, I'm cynical about
what spectra and light curves ``demand.''  These are inverse
problems, tough ones.  In any case, just because nobody is
{\bf ready} to calculate merger/core-collapse doesn't mean it 
can't or doesn't happen, and in such a way that it doesn't 
produce the few tenths of a solar mass of Si and S.  Accepting 
the "demand" of spectra and light curves leaves the ONLY answer 
to a Ia (SN 2003fg) with more than 1.2 solar masses of $^{56}$Ni 
as a "super-Chandrasekhar mass" white dwarf -- i.e., {\bf 
inventing} new physics!  That's {\bf far} worse!
 
{\it He argues that the presence of pulsars in globular clusters
suggests core collapse in Ia's, but it is possible to get
either thermonuclear explosion (Ias) or or accretion induced
collapse (pulsars) from very similar models in which only
the accretion rate is changed. Or the DD mergers may make
pulsars while another event makes SN Ia's.}
 
Always apply Occam's Razor.  Always.  Whether it's possible
to get TN explosion or CC depending on accretion rate is
still an open question.  Only models exist, and these have
been developed by those with an axe to grind.
An {\bf observation}:  the recycled pulsars in the core-collapsed
(CCd) GC Ter 5 with determined masses weighed in
at 1.7 solar -- more than 1/4 solar mass above 1.4 (Scott
Ransom 2006).  That's a LOT of mass needed to spin things up.
Thus the "superaccretion" gambit is dead.  There never was any
evidence for it in the first place.  It had been moribund ever
since my colleagues and I discovered the first pulsar in the
non-core-collapsed (nCCd) GC, M28, in 1987, and since SN 2006mr
dead and moldy.  Astronomers just keep inventing stuff so that 
they can continue to get grants to work on what they want to 
work on in a grant system that is broken through no fault of their 
own.  (``I'll support your BS if you support my BS,'' etc.)  In 
this way they become part of the problem, and it takes a good 
hard kick to bring them back to reality.  Now events in the 
universe have occurred so quickly that they can't invent things 
fast enough.  And without superaccretion, there are just TOO many 
ms pulsars in the nCCd GCs to not have most of them born fast.  
Aside from that, Middleditch, J. 2004, ApJL, 601, L167, has been 
published for three years now, and it suggests just this.  
Nothing wrong with it then.  Nothing wrong with it now.
 
{\it He argues that 87A was a DD merger of two stellar cores. 
What is that?  Does DD stand for "double degenerate"?
Sanduleak 202-69} [sic] {\it was a RSG just 20,000 years before 
it was a BSG. There are merger models for 87A, but neither of 
the merging counterparts is degenerate in any model of which I
am aware. At a minimum more details of this "model" are
needed.}
 
DD does indeed stand for double degenerate, last paragraph of
the Introduction.  Also Sanduleak -69 202.
The astrophysically negligible 10 km/s expansion of the
equatorial ring, the thermal velocity of H at 10,000 K,
is a big clue here.  Never was an RG because of overflow
out of common envelope through 1 or both outer mass-axis
Lagrangean points.  The lack of any strongly magnetized
pulsar remnant, {\bf unlike} SN 1986J, is another clue that
87A (and damn near all others) did not result from Fe
photodissociation catastrophe.  The excess of N seen
in SNe may also be a signature of merger.
 
{\it He uses the observation of polarization and high velocity
lines in SN Ia to argue for a merger, but does not present a
model to back up the claim nor consider other explanations
given in the literature (asymmetric ignition and explosion).}

It's an ApJ Letter, 4 pages TOPS!  The model is
staring us in the face as the picture of SN 1987A
(www.stsci.edu).  {\bf Better} than a model in many ways.
 
{\it He argues in a few throw away lines that a long soft GRB
results when a short hard GRB (produced how?) happens inside
a lot of matter}
 
``Throw away lines''????  This is a version of the
{\bf dreaded} "You can't just SAY that!" (YCJST) objection.
You know you've really hit home when they are desperate
enough to raise that one. People use this when it would look 
like they've been idiots in the past.  Who hasn't?  The line: 
``Chen \& Ruderman (1993[, ApJ, 408, 179]) suggested
that the formation mechanism for millisecond pulsars in
the Galactic disk and globular clusters might be 
different.'' from Camilo et al.~2000, ApJ, 535, 975, {\bf is}
a throw away line because in this particular case it, and the 
following text {\bf misrepresent} what Chen \& Ruderman said.  
Namely, that the nCCd GCs don't produce pulsars like 1920+57,
1937+21, M15 A,B,C, \& G, because these are likely to have 
been ``recycled,'' via accretion, leading to magnetic
pole migration toward the rotational axis due to accretion
stresses, and not born fast, via merger, leading to wider, more
frequently double pulse profiles.  Confirming this is the low 
luminosity of pulsars found in Ter 5, relative to pulsars
in nCCD GCs (Ransom 2006), which is very likely a selection 
effect in favor of narrow pulses because of their Fourier 
properties, making them more detectable even
than higher luminosity, higher dutycycle pulsars.
So a throw away line is a line whose flaws are easily
pointed out, which the referee has not done with the
lines he has noted, except his own throw away line 
about the ``how'' issue.
 
As for that, does anyone {\bf really} doubt that GRBs can be 
softened by transmission through matter and viewed slightly 
off axis?  In any case, witness the ``mystery spot from SN
1987A.  10$^{49}$ ergs of energy in it even 50 days after
the SN (takes only 8 extra days for a polar beam/jet to
get to it and thence on to the Earth).  We'll be figuring
this out over the next few decades.  Again, this is already
published in ApJL, 601, L167.
 
I am not {\bf inventing} anything, or claiming {\bf new physics}, 
just trying to get people to pay attention to perfectly valid 
observations and airtight logic:
                                                                               
        A.   SN 1987A tried to make a gamma-ray burst,
                (witness the mystery spot), and would
                have succeeded were it not for the Hydrogen
                and Helium sandbags in its common envelope.

        B.   No long, soft GRBs are seen in elliptical
                galaxies.

        C.   The dominant mechanism for producing objects
                more compact than white dwarfs (WDs) in
                galaxies, particularly ellipticals, is
                WD-WD merger to supernova (as occurred with
                the cores of the two stars which merged to
                produce Sk -69 202), dominated in these
                populations, as always, by binary-binary
                collisions.

        D.   But these tried to make a GRB within SN 1987A.

        E.   Therefore, WD-WD merger in ellipticals {\bf must}
                produce some kind of GRBs.

        F.   But only short, hard GRBs happen in ellipticals.

        G.   Therefore WD-WD mergers produce short, hard,
                GRBs.

{\it He states a lot of opinions that have nothing to do with
making his case - a cosmological constant is "unnatural";
people using SN Ia's to do cosmology are "unwise". He brings
back the 2 ms pulsar in 87A which he, and only he was ever
able to detect. He dismisses other observations of the CMBR
that have also provided evidence for a cosmological constant
nearly equal to what the SN Ia groups found.}
 
The CMBR first:

Yes, but they {\bf keep} saying that Ia cosmology is the {\bf 
only direct} evidence for DE (e.g., Conley et al. 2006, 
Sullivan et al. 2006).  {\bf Why} do you suppose they do that?
 
Now 87A: 

The object was found in data from many telescopes
and observatories.  Sure, I did the first pass analyses,
but reputable collaborators have also verified the signals,
and the data have been offered to {\bf anyone} who requests 
it.
 
We've even had a night in common with another group, with
an agreement to share the data (I have a {\bf slide} of the
guy with Jerry Kristian on the afternoon of 1992, Nov.~6,
in the Las Campanas 2.5-m control room).  The promised
data was never delivered, even though we did ask for it.
(WHY?  Written over inside the laptop?  Lost?  Absolute
verification of the signal too damaging to astronomers?
Ergo decades of work down the drain?)
 
I also pointed the HST/HSP collaboration to candidate
frequencies for which we found a signal in their data on
June 2, 1992, and March 6, 1993.  For whatever reason they
did not respond then, wrote a paper claiming an upper
limit of 27th magnitude, which Kristian and I refereed,
and we informed them that it was really 22 (100 times
brighter -- the HSP count rate on any object of known
magnitude will verify that the intrumental throughput
is 1\%, and from this, limits can be set from the total
number of counts in any observaion), and told ApJ 
that we'd like to see the paper again before it got 
published.  Next time we saw the paper it {\bf was} 
published with a limit of 24.5 (still exaggerated 
10 times too dim).  A representative of the collaboration 
showed up at the SN 1987A -- 10 Years After conference 
in La Serena, Chile, and tried to argue for this 
limit, at least until he showed a power spectrum of his 
calibration object.  When I informed him that the object 
had 10 times less background than SN 1987A, and was also 
integrated over a 50\% longer time interval, he could only 
leave the stage, muttering.
 
Manchester and Peterson published on not seeing a signal
in Dec. of `94 (I looked at their data, they {\bf really} 
didn't see anything.)  But they spent only part of the two 
nights on 87A.  In fact, at the Aspen SN 1987 \& GRBs
Conference during Feb. 19-23), Manchester made a whole 
contributed talk on the basis of this observation, plus
the published times 10 exaggerated faint limit (24.5) from 
the HSP.  I had to correct that hearsay claim on the
spot.

Aside from that, it's not like there {\bf 
aren't} 10 solar masses of starguts moving around,  or any 
pulsar remnant isn't precessing and potentially changing its 
beaming.  As far as I know, they had no observations during 
the interval from Feb. of `92 through Sep. of `93 (they tried 
on September 15, 1993, but were clouded out -- signals 
were seen from Tasmania on the 12th and 24th), when we were 
detecting the signal most consistently.  Remember, HST was 
still nearsighted during that interval.

So THAT was our competition!
                                                                               
{\it In short, if he had taken any single one of his claims,
e.g., that the SN Ia sample might be biased by including
some core collapse events along with the rest, and treated
that claim carefully, maybe even joined with a model builder
to flesh out the claim, a worthy publication might have
resulted.}
                                                                               
Again:  It's an ApJ Letter, 4 pages tops.  See above for
ApJL policy.

\end{document}